\documentclass{aastex61}

\usepackage[version=4]{mhchem}

\received{}
\revised{}
\accepted{}
\submitjournal{ApJ}

\shorttitle{Depletion of heavy nitrogen}
\shortauthors{Furuya \& Aikawa}

\begin{document}

\title{Depletion of heavy nitrogen in the cold gas of star-forming regions}

\correspondingauthor{Kenji Furuya}
\email{furuya@ccs.tsukuba.ac.jp}

\author{Kenji Furuya}
\affiliation{Center for Computer Sciences, University of Tsukuba, 305-8577 Tsukuba, Japan}
\author{Yuri Aikawa}
\affiliation{Department of Astronomy, Graduate School of Science, The University of Tokyo, 7-3-1 Hongo, Bunkyo-ku, Tokyo 113-0033, Japan}



\begin{abstract}
We investigate nitrogen isotope fractionation in forming and evolving molecular clouds using gas-ice astrochemical simulations.
We find that the bulk gas can become depleted in heavy nitrogen ($^{15}$N) due to the formation of $^{15}$N-enriched ices.
Around the chemical transition from atomic nitrogen to \ce{N2}, N$^{15}$N is selectively photodissociated, which results in the 
enrichment of $^{15}$N in atomic nitrogen.
As $^{15}$N-enriched atomic nitrogen is converted to ammonia ice via grain surface reactions, the bulk gas is depleted in $^{15}$N.
The level of $^{15}$N depletion in the bulk gas can be up to a factor of two compared to the elemental nitrogen isotope ratio, 
depending on the photodesorption yield of ammonia ice.
Once the nitrogen isotopes are differentially partitioned between gas and solids in a molecular cloud, 
it should remain in the later stages of star formation (e.g., prestellar core) as long as the sublimation of ammonia ice is inefficient.
Our model suggests that all the N-bearing molecules in the cold gas of star-forming regions can be depleted in $^{15}$N, 
which is at least qualitatively consistent with the observations toward prestellar core L1544.
In our models, icy species show both $^{15}$N and deuterium fractionation.
The fractionation pattern within ice mantles is different between $^{15}$N and deuterium, reflecting their fractionation mechanisms;
while the concentration of deuterium almost monotonically increases from the lower layers of the ice mantles to the upper layers, 
the concentration of $^{15}$N reaches the maximum at a certain depth and declines towards the surface.

\end{abstract}

\keywords{astrochemistry --- ISM: molecules --- ISM: clouds --- stars: formation}



\section{Introduction} \label{sec:intro}
Molecular isotope ratios are essential tools to investigate the origin of solar system materials and their possible chemical link with interstellar materials.
Nitrogen has two isotopes, $^{14}$N and $^{15}$N.
The most primitive materials in the solar system, such as cometary ices (e.g., HCN and \ce{NH3}), show the enrichment of $^{15}$N by a factor of up to several 
compared to the Sun \citep[e.g.,][]{marty11,mumma11,shinnaka16}.
There remain open questions as to what the main cause of $^{15}$N fractionation in the solar system materials was, and when and where the fractionation was implemented.
To address these questions, studies of $^{15}$N fractionation in star- and planet-forming regions are crucial.

The elemental abundance ratio [$^{14}$N/$^{15}$N]$_{\rm elem}$ in the local ISM  has been estimated to be $\sim$200-300 
from observations toward diffuse clouds. 
\citet{ritchey15} estimated [$^{14}$N/$^{15}$N]$_{\rm elem} = 274 \pm 18$ from the studies of CN absorption lines in the optical,
while \citet{lucas98} estimated [$^{14}$N/$^{15}$N]$_{\rm elem} = 237^{+27}_{-21}$ from the studies of HCN absorption lines in the mm range.
In diffuse clouds, nitrogen isotope fractionation would be efficient neither by isotope exchange reactions nor by isotope selective photodissociation of \ce{N2} (see below), 
given that the gas temperature is warm ($\gtrsim$20 K) and the majority of elemental nitrogen is likely present in the atomic form \citep[e.g.,][]{knauth04,ritchey15}.
The estimated [$^{14}$N/$^{15}$N]$_{\rm elem}$ ratio in the local ISM is lower than that in the Sun \citep[441,][]{marty11}, 
indicating the galactic chemical evolution over the last 4.6 billion years \citep[e.g.,][]{romano17}.

Recent observations have quantified the degree of $^{15}$N fractionation in various molecules in low-mass dense cores \citep[e.g.,][]{bizzocchi13,daniel13,daniel16,gerin09,hilyblant13a,hilyblant13b}.
The measurements in L1544 (prestellar core) and B1 \citep[a core possibly harboring the first hydrostatic core; e.g.,][]{pezzuto12} are summarized in Table \ref{table:obs}. 
Among the measurements, the depletion of $^{15}$N in \ce{N2H+} is the most puzzling, 
because its parent molecule, \ce{N2}, may be the primary reservoir of gaseous nitrogen and because astrochemical models have failed to explain this trend as discussed below.
In particular, in L1544 \ce{N2H+} is depleted in $^{15}$N by a factor of around four compared to [$^{14}$N/$^{15}$N]$_{\rm elem}$ in the local ISM \citep{bizzocchi13}.

\begin{table}
\begin{center}
\caption{Observed and model N/$^{15}$N ratios \label{table:obs}}
\begin{tabular}{cccccc}
\hline\hline
                                            & L1544\tablenotemark{a),}\tablenotemark{b),}\tablenotemark{c),}\tablenotemark{d)}   & B1\tablenotemark{e)}  & \multicolumn{3}{c}{Model\tablenotemark{f)}}   \\ 
\cline{4-6}
                                            &                          &          & $Y_{\rm pd}^{\rm NH_3}=5\times10^{-4}$   & $Y_{\rm pd}^{\rm NH_3}=10^{-3}$ & $Y_{\rm pd}^{\rm NH_3}=3\times10^{-3}$  \\
\hline
\ce{N2H+}/N$^{15}$NH$^+$       & $1050 \pm 220$   & 400$^{+100}_{-65}$       & 524 (517,518) & 410 (405,406)     & 313 (308,310)          \\
\ce{N2H+}/$^{15}$NNH$^+$       & $1110 \pm 240$   & $>$600                      & 524 (517,518) & 411 (405,407)    & 315 (312,310)         \\
\ce{NH3}/$^{15}$\ce{NH3}         &  --                    & 300$^{+55}_{-40}$        & 372 (447,492) & 400 (371,388)      & 374 (290,297)          \\
\ce{NH2D}/$^{15}$\ce{NH2D}     & $>$700               & 230$^{+105}_{-55}$       & 371 (448,492) & 384 (371,388)     & 349 (289,297)          \\
CN/C$^{15}$N\tablenotemark{g)} & $500 \pm 75$      & 290$^{+160}_{-80}$       & 335  (383,501) & 292 (338,383)     & 224  (257,276)         \\
HCN/HC$^{15}$N\tablenotemark{g)} & 257                & 330$^{+60}_{-50}$         & 342 (390,502) & 306  (349,385)    & 244  (271,280)         \\
N/$^{15}$N (bulk gas)               &  --                     & --                             & 495 (513,518) & 410 (408,407)      & 320 (314,311)         \\
\ce{NH3}/$^{15}$\ce{NH3} (ice)  & --                & --                             & 292 (292,282)  &  272 (274,276)    & 246 (252,259)          \\
\ce{HCN}/HC$^{15}$N (ice)       & --                      & --                        &  274 (293,326)  &  238 (276,283)   & 193 (196,220)          \\
\hline
\end{tabular}
\end{center}
\tablecomments{$^{a)}$\citet{bizzocchi13}\\
$^{b)}$\citet{gerin09}\\
$^{c)}$\citet{hilyblant13a}\\
$^{d)}$\citet{hilyblant13b}\\
$^{e)}$\citet{daniel13}\\
$^{f)}Y_{\rm pd}^{\rm NH_3}$ is the photodesorption yield of \ce{NH3}.
The values are the results at $A_V = 3$ mag, while 
the first (second) values in parentheses are the ratios after the additional 10$^5$ (10$^6$) yr evolution under prestellar core conditions.\\
$^{g)}$The ratios were derived from the observations of the $^{13}$CN/C$^{15}$N ratio and the H$^{13}$CN/HC$^{15}$N ratio, 
assuming the elemental C/$^{13}$C ratio.
}
\end{table}

The chemical network of nitrogen isotope fractionation triggered by isotope exchange reactions was proposed by \citet{terzieva00} and updated by \citet{roueff15}.
Following these works, there have been many numerical studies on nitrogen isotope fractionation in molecular clouds and prestellar cores in the framework of
pseudo-time-dependent models \citep[e.g.,][]{charnley02,rodgers08a,wirstrom12,hilyblant13b,roueff15,wirstrom17}.
These models are diverse in terms of the primary nitrogen reservoir in the gas phase (\mbox{\ion{N}{1}} or \ce{N2}), 
the degree of CO freeze out (from non-depleted to fully depleted), the adopted chemical network (e.g., with or without nuclear spin state chemistry of \ce{H2}), 
and physical conditions (density of 10$^{4-6}$ cm$^{-3}$).
Neverthless, common qualitative prediction is that \mbox{\ion{N}{1}} is depleted in $^{15}$N, while \ce{N2} is enriched in $^{15}$N.
This is a natural consequence of the lower zero point energy of N$^{15}$N compared to $^{14}$\ce{N2} \citep[e.g.,][]{roueff15}.
Since \ce{N2H+} forms from \ce{N2}, the previous models have predicted that \ce{N2H+} is enriched in $^{15}$N, which contradicts the observations \citep{wirstrom12}.
Moreover, \citet{roueff15} have shown that the $^{15}$N fractionation triggered by isotope exchange reactions is much less effective than had been
previously thought, due to the presence of activation barriers for some key reactions, such as $^{15}$N + \ce{N2H+}.

On the other hand, $^{15}$N depletion in \ce{N2H+} could be explained by isotope selective photodissociation of \ce{N2} \citep{liang07,heays14}.
Photodissociation of \ce{N2} is subject to self-shielding \citep{li13}.
Because N$^{15}$N is much less abundant than \ce{N2}, N$^{15}$N needs a higher column density of the ISM gas for self-shielding than $^{14}$\ce{N2}.
As a result, in some regions, N$^{15}$N is selectively photodissociated with respect to \ce{N2}, and then \ce{N2} is depleted in $^{15}$N, 
while the photofragment, \mbox{\ion{N}{1}}, is enriched in $^{15}$N.
\citet{heays14} developed a depth-dependent pseudo-time-dependent model of a molecular cloud, considering  
both isotope selective photodissociation of \ce{N2} and a set of nitrogen isotope exchange reactions.
They found that the isotope selective photodissociation is at work and \ce{N2} (and \ce{N2H+}) are depleted in $^{15}$N, 
but only in the chemical transition zone from \mbox{\ion{N}{1}} to \ce{N2}, where the interstellar FUV radiation field is not significantly attenuated, 
i.e., visual extinction of  a few mag.
Prestellar cores, however, typically have higher visual extinction ($>$10 mag for a dust continuum peak) compared to their cloud model.

In this paper, we propose that nitrogen isotope fractionation in prestellar cores are largely inherited from their parent clouds, where interstellar UV radiation can penetrate,
based on our physico-chemical models.
We investigate the nitrogen isotope fractionation in a forming and evolving molecular cloud via converging flow.
As in \citet{heays14}, \mbox{\ion{N}{1}} is enriched in $^{15}$N at the \mbox{\ion{N}{1}}/\ce{N2} chemical transition via isotope selective photodissociation of \ce{N2}. 
In our models, $^{15}$N-enriched \mbox{\ion{N}{1}} is frozen out onto grains and converted to \ce{NH3} ice, which depletes $^{15}$N from the bulk gas.
Once the nitrogen isotopes are differentially partitioned making $^{15}$N-depleted gas and $^{15}$N-enriched solids in a molecular cloud, 
it should remain in the later stages of star formation (e.g., prestellar core) as long as dust temperature is cold and ice sublimation is inefficient.
This process can be seen as an analog of oxygen isotopic anomaly production in the solar system materials through isotope selective photodissociation 
of CO \citep{clayton02,yurimoto04,furi15}, while, to the best of our knowledge, it has never been investigated in details. 

This paper is organized as follows.
We describe our physical and chemical models in Section \ref{sec:model}.
In Section \ref{sec:result}, we show that the bulk gas becomes depleted in $^{15}$N in our models and discuss the mechanism.
In Section \ref{sec:discussion}, we extend our model to the prestellar core phase, which is compared to the prestellar core observations.
The nitrogen and dueterium isotope fractionation of icy species are also briefly discussed.
We conclude this paper in Section \ref{sec:conclusion}.

\section{Model} \label{sec:model}
We simulate molecular evolution in a forming and evolving molecular cloud.
Since our physical and chemical models are similar to those in \citet{furuya15} and \citet{furuya18} except for the inclusion of $^{15}$N fractionation chemistry,
we present here the brief summary of the models.

\subsection{Physical model}
One of the plausible scenarios of molecular cloud formation is that diffuse \mbox{\ion{H}{1}} gas is compressed by super-sonic accretion flows \citep[e.g.,][]{inoue12}.
To simulate the physical evolution of post-shock materials (i.e., forming cloud),  we use the steady shock model developed 
by \citet{bergin04} and \citet{hassel10}.
The model solves the conservation laws of mass and momentum with the energy equation, 
considering time-dependent cooling/heating rates and simplified chemistry in a plane parallel configuration.
The cloud is assumed to be irradiated by the Draine field \citep{draine78} and the cosmic ray ionization rate is set to be $1.3\times10^{-17}$ s$^{-1}$.
As time proceeds, the column density of post-shock materials increases, which assists molecular formation by attenuating the UV radiation.
The column density of post-shock materials at a given time $t$ after
passing through the shock front is $N_{\rm H} \approx 2\times10^{21}~(n_0/10~{\rm cm^{-3}})(v_0/15~{\rm km s^{-1}})(t/4~{\rm Myr})~{\rm cm^{-2}}$, 
where $n_0$ and $v_0$ are preshock \mbox{\ion{H}{1}} gas density and velocity of the accretion flow, respectively.
We convert $N_{\rm H}$ into visual extinction, $A_V$, by the formula $A_V/N_{\rm H} = 5\times 10^{-22}~{\rm mag/cm^{-2}}$.
The simulation is performed until $A_V$ reaches 3 mag (i.e., $\sim$12 Myr).
In the most of the simulation time, the density and temperature of the cloud is $\sim$10$^4$ cm$^{-3}$ and 10-15 K, respectively \citep[see Fig. 2 of][]{furuya15}.

\subsection{Chemical model}
Our astrochemical model takes into account gas-phase chemistry, interactions between gas and (icy) grain surfaces, and grain surface chemistry.
We adopt a three phase model \citep{hasegawa93b}, assuming the top four monolayers are chemically active; 
the rest of the ice mantles are assumed to be inert.
Our chemical network is based on the gas-ice network of \citet{garrod06} and has been extended to include 
deuterium chemistry (up to triply deuterated species) and nuclear spin state chemistry of \ce{H2}, \ce{H3+}, and their isotopologues.
For this work, we excluded species containing chlorine, phosphorus, or more than three carbon atoms in order to shorten the computational time.
In our model, the ortho-to-para ratio of \ce{H2} is mostly determined by the competition between two processes; 
the ratio is three upon the formation of \ce{H2} on grain surfaces \citep[e.g.,][]{watanabe10}, 
while the ratio is reduced via the proton exchange reactions between \ce{H2} and \ce{H+} (or \ce{H3+}) \citep[e.g.,][]{honvault11}.
The reaction rate coefficients of the \ce{H2}-\ce{H3+} system is taken from \citet{hugo09}.
Gas-phase nitrogen chemistry has been updated referring to \citet{wakelam13} and \citet{loison14}.
For this work, the chemical network is extended to include mono-$^{15}$N species, relevant isotope exchange reactions \citep{roueff15}, 
and isotope selective photodissociation of \ce{N2} \citep{li13,heays14}.
We therefore added 460 species and $\sim$26,000 reactions to our reduced deuterated network, resulting in the increase of the number of 
species from 1350 to 1810 and the number of reactions from $\sim$73,000 to $\sim$99,000.

The binding energy of each species is not constant in our simulations,
but depends on the composition of an icy surface, following the method presented in \citet{furuya18}.
The binding energy of species $i$, $E_{\rm des}(i)$, is calculated as a function of surface coverage of species $j$, $\theta_j$, where $j$ = \ce{H2}, CO, \ce{CO2}, or \ce{CH3OH}:
\begin{equation}
E_{\rm des}(i) = (1-\Sigma_j\theta_j)E^{\ce{H2O}}_{\rm des}(i) + \Sigma_j\theta_j E^{j}_{\rm des}(i), \label{eq:edes}
\end{equation}
where $E^j_{\rm des}(i)$ is the binding energy of species $i$ on species $j$.
The set of the binding energies on water ice, $E^{\ce{H2O}}_{\rm des}$, is taken from \citet{garrod06} and \citet{wakelam17}.
In particular for this study, $E^{\ce{H2O}}_{\rm des}$ are set to be 550 K for \mbox{\ion{H}{1}} and \ce{H2}, 720 K for \mbox{\ion{N}{1}}, 1170 K for \ce{N2}, and 5500 K for \ce{NH3} \citep{collings04,fayolle16,minissale16}.
There is no laboratory data or estimate for most $E^{j}_{\rm des}(i)$ in the literature.
In order to deduce $E^{j}_{\rm des}$ for all species, where $j$ is either \ce{H2}, CO, \ce{CO2}, or \ce{CH3OH}, we assume scaling relations \citep[cf.][]{taquet14},
\begin{align}
E^{j}_{\rm des}(i) = \epsilon_j E^{\ce{H2O}}_{\rm des}(i),
\end{align}
where $\epsilon_j$ is $E^{j}_{\rm des}(j)/E^{\ce{H2O}}_{\rm des}(j)$.
We adopt $\epsilon_{\ce{H2}}$ = 23/550, $\epsilon_{\ce{CO}}$ = 855/1300, $\epsilon_{\ce{CO2}}$ = 2300/2690, and $\epsilon_{\ce{CH3OH}}$ = 4200/5500 \citep[e.g.,][]{oberg05,cuppen07,noble12}.
For example, the binding energy of \ce{N2} becomes 770 K in our simulations when an icy surface is fully covered by CO \citep{oberg05,fayolle16}.
For sticking probabilities of gaseous species onto (icy) grain surfaces, we use the formula recommended by \citet[][their Eq. 1]{he16}, 
which leads to the sticking probability of around unity for all species for the relevant temperature range in our physical model.
For $^{15}$N- and/or D-bearing species, we use the same binding energies and sticking probabilities as for normal species.
The exception is the binding energy of atomic deuterium, whose binding energy is set to be 21 K higher than that of atomic hydrogen \citep{caselli02}.
Adsorption rates are inversely proportional to the square root of the mass of species, 
so that adsorption of gaseous species tends to make the bulk gas enriched in heavier isotopes, D and $^{15}$N.
This effect is not important especially for $^{15}$N, because the difference in the rates is only a few \%.

As non-thermal desorption processes, we consider stochastic heating by cosmic-rays, photodesorption, and chemical desorption \citep[e.g.,][]{hasegawa93a,westley95,garrod07}.
We assume that species formed by surface reactions are desorbed by chemical desorption with the probability of roughly 1 \% \citep{garrod07}.
Photodissociation rates of icy species are calculated in the same way as \citet{furuya15}.
Photodissociation occurs only in the surface layers (i.e., top four monolayers) in our model, since the rest of ice mantle is assumed to be chemically inert.
In other words, photofragments is assumed to immediately recombine in the bulk ice mantle in our model \citep{furuya17}.
According to molecular dynamics (MD) simulations, there are several possible outcomes after a UV photon dissociates water ice; 
(i) the photofragments are trapped on the surface; 
(ii) either of the fragments is desorbed into the gas phase; 
(iii) the fragments recombine and the product is either trapped on the surface or desorbed into the gas phase, etc. 
\citep[][; see also e.g., Hama et al. (2010) for experimental studies]{andersson08,arasa15}.
Note that desorption of the photofragments or the recombination product occurs from the top several monolayers only \citep{andersson08,arasa15}.
To the best of our knowledge, similar MD simulations for other molecules have not been reported in the literature.
For water ice photodissociation, we consider all the possible outcomes and their branching ratios 
found by the MD simulations of \citet{arasa15}.
For photodissociation of other icy species, we assume the same branching ratios as in the gas-phase photodissociation, 
and that all the photofragments are trapped on the surface for simplicity.
For those species, photodesorption rate is given as following, separately from photodissociation rate:
\begin{align}
R_{{\rm phdes}, \,i} = \pi a^2 n_{\rm gr} [F_{\rm ISUV} \exp(-\gamma_i A_V) + F_{\rm CRUV}] 
\times \theta_i Y_i{\rm min}(N_{\rm layer}/4, \,\, 1),
\end{align}
where $n_{\rm gr}$ is the number density of dust grains, 
$a$ is the radius of dust grains (0.1 $\mu$m), 
$F_{\rm ISUV}$ and $F_{\rm CRUV}$ are the interstellar and cosmic-ray induced FUV photon fluxes, respectively,
$\gamma_i$ is the parameter for the attenuation of interstellar radiation field by dust grains,
$Y_i$ is the ptotodesorption yield per incident FUV photon for thick ice ($\geq$4 monolayers),
and $N_{\rm layer}$ is the number of monolayers of the whole ice mantle.
The parameter $\theta_i$ is the surface coverage of species $i$,
which is defined as a fractional abundance of species $i$ in the top four monolayers.
The photodesorption yield for \ce{N2} ice per incident FUV photon is given as an increasing function of the surface coverage of CO, 
varying from $3\times10^{-3}$ to $8\times10^{-3}$ \citep{bertin13}.
We assume that \ce{NH3} ice is photodesorbed intact (as \ce{NH3}) with the yield of $10^{-3}$.
This value is similar to the recently measured photodesorption yield for pure \ce{NH3} ice in laboratory \citep[$2.1^{+2.1}_{-1.0}\times10^{-3}$,][]{martin17}, 
but the yield would depend on ice compositions and the FUV spectrum adopted in experiments \citep{ligterink15,martin17}.
The dependence on the \ce{NH3} photodesorption yield is discussed in Section \ref{sec:photodes}.
We assume the same photodesorption yields for $^{15}$N- and/or D-bearing species as for normal species.

Elemental abundance ratios for H:He:C:N:O:Na:Mg:Si:S:Fe are 
1.00:9.75(-2):7.86(-5):2.47(-5):1.80(-4):2.25(-9):1.09(-8):9.74(-9):9.14(-8):2.74(-9), 
where $a(-b)$ means $a\times10^{-b}$  \citep{aikawa99}.
Elemental abundance ratios, [D/H]$_{\rm elem}$ and [N/$^{15}$N]$_{\rm elem}$, are set to be $1.5\times10^{-5}$ and 300, respectively \citep{linsky03,ritchey15}.
All the elements, including hydrogen, are initially in the form of either neutral atoms or atomic ions, depending on their ionization energy.

\section{Result} \label{sec:result}
\subsection{Fiducial model}
\begin{figure}[ht!]
\plotone{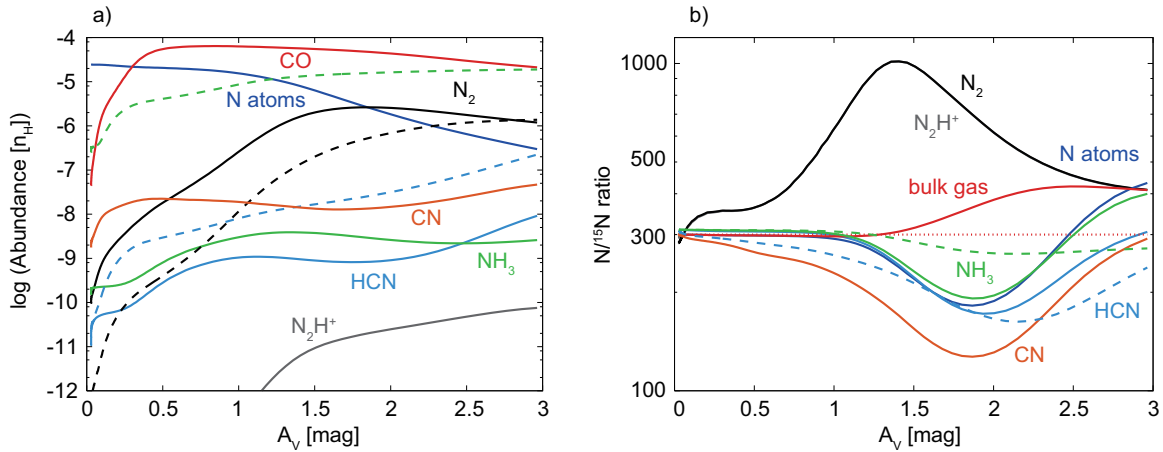}
\caption{Molecular abundances of selected species (panel a) and N/$^{15}$N ratios (panel b) as functions of $A_V$.
Solid lines indicate gaseous species, while the dashed lines indicate species in the whole ice mantle.
The red dotted line in panel (b) indicates [N/$^{15}$N]$_{\rm elem} = 300$,
while the red solid line indicates the N/$^{15}$N ratio in the bulk gas.
Note that the N/$^{15}$N ratio of \ce{N2} is twice of the \ce{N2}/N$^{15}$N abundance ratio.
The N/$^{15}$N ratios of \ce{N2} and \ce{N2H+} are almost identical and indistinguishable.
\label{fig:n_ab}
}
\end{figure}

Figure \ref{fig:n_ab}(a) shows abundances of N-bearing species as functions of $A_V$.
With increasing $A_V$, \mbox{\ion{N}{1}} is gradually converted to \ce{NH3} ice and \ce{N2}.
At the final simulation time in our model ($\sim$12 Myr), which corresponds to $A_V = 3$ mag, most nitrogen is locked in molecules, 
and the fraction of elemental nitrogen in \mbox{\ion{N}{1}}, \ce{N2} (gas+ice), and \ce{NH3} ice are
0.05 \%, 20 \% and 80 \%, respectively.
Note that $\sim$90 \% of nitrogen is in ice at $A_V = 3$ mag.
In our models, the partitioning of elemental nitrogen among the three species depends on the photodesorption yield of ammonia ice
as shown in \citet{furuya18} and briefly discussed in Section \ref{sec:photodes}.
Observationally, the primary nitrogen reservoir in dense star forming regions is not well constrained, 
although it has been expected to be either \mbox{\ion{N}{1}}, gaseous \ce{N2}, or icy N-bearing species, such as \ce{N2} and \ce{NH3} \citep[e.g.,][]{maret06,daranlot12}.

The N/$^{15}$N abundance ratios in selected species are shown in Figure \ref{fig:n_ab}(b).
In our model, $^{15}$N fractionation is predominantly caused by isotope selective photodissociation of \ce{N2} rather than isotope exchange reactions.
At $A_V \sim 1.8$ mag, the primary reservoir of nitrogen in the gas phase changes from \mbox{\ion{N}{1}} to \ce{N2}.
Around this \mbox{\ion{N}{1}}/\ce{N2} transition, \mbox{\ion{N}{1}} is significantly enriched in $^{15}$N (a factor of up to two compared to [N/$^{15}$N]$_{\rm elem}$), 
while \ce{N2} is significantly depleted in $^{15}$N.
Isotope selective photodissociation of \ce{N2} is at work only in the limited regions ($1\,\,{\rm mag} \lesssim A_V \lesssim 2.5\,\, {\rm mag}$), 
where the interstellar UV radiation field is not significantly attenuated \citep[see also][]{heays14}.
On the other hand, it is seen that the N/$^{15}$N ratio of the bulk gas starts to increase around the \mbox{\ion{N}{1}}/\ce{N2} transition.
This is due to the freeze out of $^{15}$N-enriched \mbox{\ion{N}{1}} and subsequent \ce{NH3} ice formation, which result in the $^{15}$N depletion in the bulk gas.
Even at $A_V \sim 3$ mag, where \ce{N2} photodissociation is negligible, the bulk gas is depleted in $^{15}$N with the N/$^{15}$N ratio of $\sim$400.
The freeze-out of \ce{N2} is less efficient than that of \mbox{\ion{N}{1}} in our fiducial model, 
because adsorbed \ce{N2} does not react with other species in contrast to \mbox{\ion{N}{1}}, 
and adsorbed \ce{N2} partly goes back to the gas phase via photodesorption and stochastic heating by cosmic rays.
The binding energy of \ce{N2} is much lower than that of \ce{NH3}, while the photodesorption yield of \ce{N2} is higher than that of \ce{NH3} in our fiducial model 
($>$3$\times10^{-3}$ versus 10$^{-3}$).
Then most nitrogen lost from the gas phase is in the form of \mbox{\ion{N}{1}} rather than \ce{N2} around the \mbox{\ion{N}{1}}/\ce{N2} transition in our fiducial model, 
leading to the $^{15}$N depletion in the bulk gas.

\subsection{Dependence on the ptotodesorption yield of \ce{NH3} ice}
\label{sec:photodes}
The degree of $^{15}$N depletion in the bulk gas depends on how much of $^{15}$\mbox{\ion{N}{1}} is frozen out and 
converted to $^{15}$\ce{NH3} ice around the \mbox{\ion{N}{1}}/\ce{N2} transition.
The dominant destruction process of \ce{NH3} ice in our fiducial model is photodesorption, assuming the yield of 10$^{-3}$ per incident FUV photon.
It is expected that the degree of $^{15}$N depletion in the bulk gas depends on the \ce{NH3} photodesorption yield ($Y_{\rm pd}^{\rm NH_3}$).
Note that photodissociation of \ce{NH3} ice, the products of which are assumed to be \ce{NH2}$_{\rm ice}$ + H$_{\rm ice}$ or \ce{NH}$_{\rm ice}$ + \ce{H2}$_{\rm ice}$, 
is included in our model with much higher rate than its photodesorption (by a factor of $>$10).
A significant fraction of the photofragments, however, is hydrogenated to reform \ce{NH3} ice.
The late limiting step of \ce{NH3} ice formation is adsorption of \mbox{\ion{N}{1}} onto grain surfaces in our model, 
and the reformation of \ce{NH3} ice via surface hydrogenation reactions is efficient enough to compensate for its photodissociation. 
Then the dominant destruction process of \ce{NH3} ice is photodesorption rather than photodissociation.
In our fiducial model, the rates of photodissociation and photodesorption of \ce{NH3} ice are calculated separately, 
while the MD simulations have shown that, at least for water ice, photodesorption is some of the possible outcomes 
of photodissociation \citep[e.g.,][]{andersson08}.
In order to investigate the dependence of our results on the assumptions on photodissociation and photodesorption of \ce{NH3} ice,
we ran an additional model, in which photodissociation of \ce{NH}$_{\rm n}$ ice (n = 1, 2, 3) leads to several outcomes, 
including desorption of \ce{NH}$_{\rm n}$ or \ce{NH}$_{\rm (n-1)}$, assuming that the probability of each outcome is the same as 
that of water ice photodissociation obtained by the MD simulations of \citet{arasa15}.
The most significant difference between this modified model and our fiducial model is that in the former, photodissociation of \ce{NH}$_{\rm n}$ ice most likely 
leads to \ce{NH}$_{\rm (n-1)ice}$ + H$_{\rm gas}$ rather than \ce{NH}$_{\rm (n-1)ice}$ + H$_{\rm ice}$.
In the modified model, the total photodesorption yield of \ce{NH3} ice (desorbed as \ce{NH2} or \ce{NH3}) per incident photon is $\sim6\times10^{-4}$.
We confirmed that our result is robust; the bulk gas is depleted in $^{15}$N with the N/$^{15}$N ratio of 440 at $A_V = 3$ mag in the modified model.
The evolution of the abundances and the N/$^{15}$N ratios of the major nitrogen species is also similar in the fiducial and modified models.
In the rest of this paper, we use our original treatment of photodissociation and photodesorption of \ce{NH3} ice, but varying $Y_{\rm pd}^{\rm NH_3}$.

\begin{figure}[ht!]
\plotone{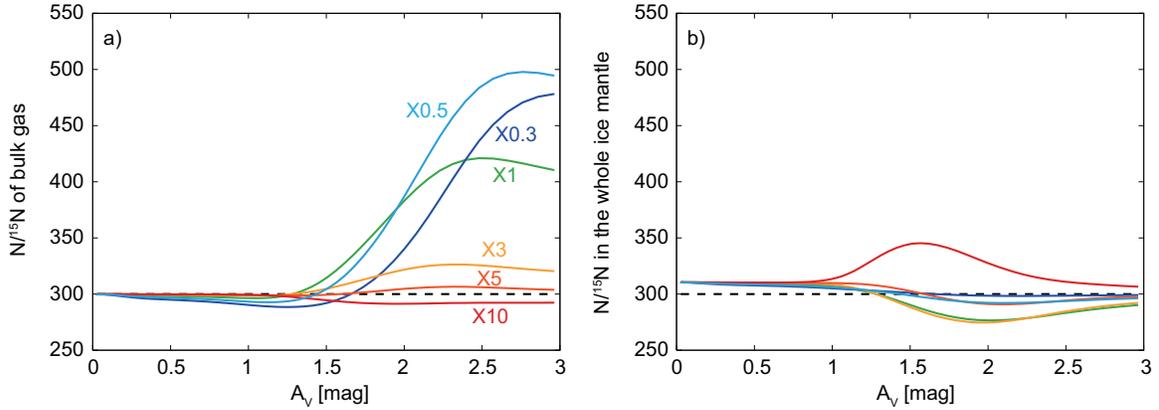}
\caption{The N/$^{15}$N ratio in the bulk gas (a) and in the bulk ice (b) as functions of $A_V$, varying the \ce{NH3} photodesorption yield 
from $3\times10^{-4}$ (blue) to 10$^{-2}$ (red). The line labeled ``$\times a$'' indicates the model with the yield of $a\times 10^{-3}$.
\label{fig:15N_gas}
}
\end{figure}

Figure \ref{fig:15N_gas} depicts variations of the N/$^{15}$N ratio in the bulk gas (panel a) and in the bulk ice (panel b) when $Y_{\rm pd}^{\rm NH_3}$ is varied 
in the range between $3\times10^{-4}$ and $10^{-2}$.
In general, the model with lower $Y_{\rm pd}^{\rm NH_3}$ shows the larger level of $^{15}$N depletion in the bulk gas.
The bulk gas $^{15}$N depletion is most significant in the model with $Y_{\rm pd}^{\rm NH_3} = 5\times10^{-4}$;
N/$^{15}$N ratio is $\sim$500, which is comparable to some of the observed molecular N/$^{15}$N ratios (see Section. \ref{sec:obs}).
Compared to the degree of the bulk gas $^{15}$N depletion, the enrichment of $^{15}$N in the bulk ice is modest, because more nitrogen is present in the ice than in the gas phase.

\begin{figure}[ht!]
\plotone{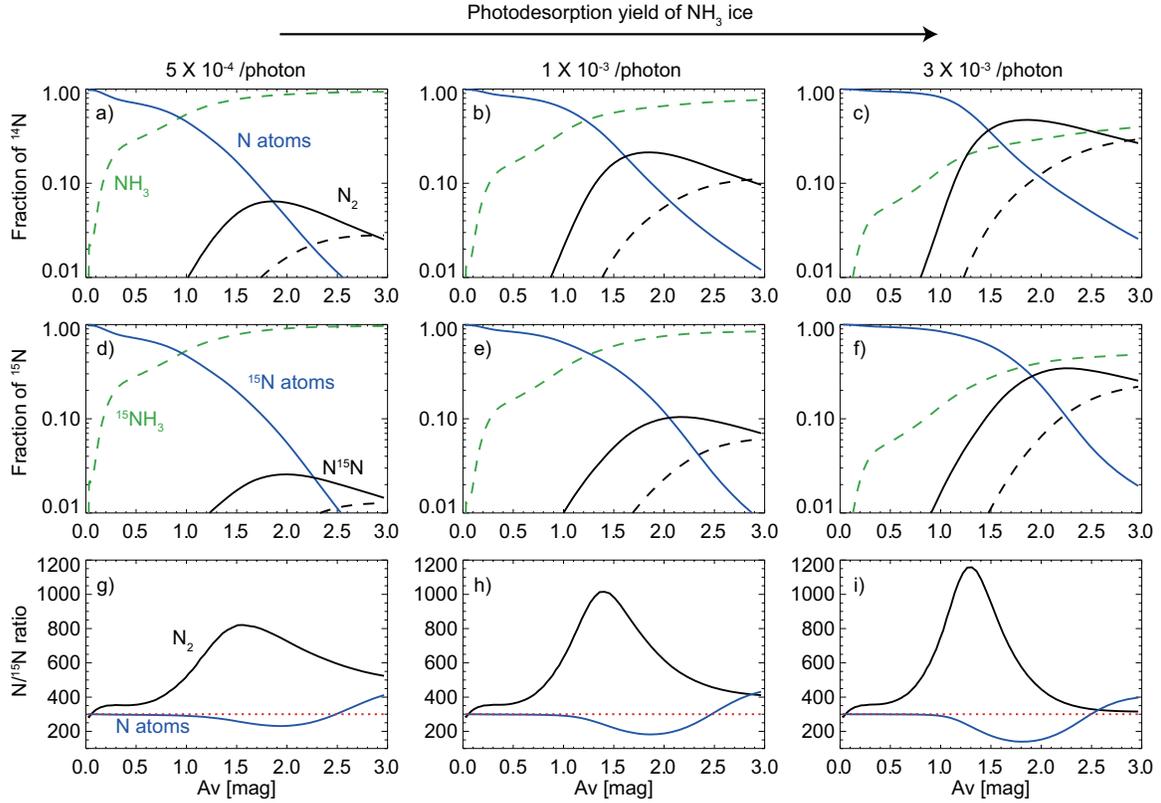}
\caption{Panels a,b,c,d,e,f) A fraction of elemental $^{14}$N (a,b,c) and $^{15}$N (d,e,f) in the forms of atomic nitrogen, \ce{N2}, and \ce{NH3} as functions of $A_V$. 
Solid lines indicate gaseous species, while the dashed lines indicate icy species.
From left panels to right panels, the \ce{NH3} photodesorption yield increases from $5\times10^{-4}$ to $3\times10^{-3}$.
Panels g,h,i) The N/$^{15}$N ratios in atomic nitrogen and \ce{N2}.
\label{fig:n_budget}
}
\end{figure}

Figure \ref{fig:n_budget} shows fractions of elemental $^{14}$N (panels a,b,c) and $^{15}$N (d,e,f) locked in \mbox{\ion{N}{1}}, \ce{N2}, and \ce{NH3} as functions of $A_V$ 
in the models with  $Y_{\rm pd}^{\rm NH_3} = 5\times10^{-4}$ (left panels), $10^{-3}$ (middle, our fiducial model), and $3\times10^{-3}$ (right).
With increasing $Y_{\rm pd}^{\rm NH_3}$, the fraction of nitrogen in \ce{NH3} ice is reduced, while that in \ce{N2} is enhanced \citep{furuya18}.
In the models with higher $Y_{\rm pd}^{\rm NH_3}$, the self-shielding of \ce{N2} is more efficient, which makes the \mbox{\ion{N}{1}}/\ce{N2} transition sharper, 
and enhances the maximum level of $^{15}$N enrichment in \mbox{\ion{N}{1}} (Fig. \ref{fig:n_budget}, panels g,h,i).
Neverthless, with increasing $Y_{\rm pd}^{\rm NH_3}$ the level of the bulk gas $^{15}$N depletion is lowered as discussed above.
The $^{15}$N fractionation between the gas and the ice mostly occurs between the $^{14}$\mbox{\ion{N}{1}}/$^{14}$\ce{N2} transition 
and the $^{15}$\mbox{\ion{N}{1}}/$^{14}$N$^{15}$N transition.
With increasing $Y_{\rm pd}^{\rm NH_3}$, the accumulation of \ce{NH3} ice is slowed down, 
while the $^{14}$\ce{N2}/$^{14}$\mbox{\ion{N}{1}} abundance ratio in the gas phase becomes higher between the two transitions.
Then the freeze out of $^{15}$N-depleted \ce{N2} becomes more important with increasing $Y_{\rm pd}^{\rm NH_3}$, 
which (partly) cancels out the bulk gas $^{15}$N depletion through the freeze out of $^{15}$N-enriched \mbox{\ion{N}{1}}.
In the model with $Y_{\rm pd}^{\rm NH_3} = 10^{-2}$, where the photodesorption yield of \ce{NH3} ice is higher than that of \ce{N2} ice, 
the bulk gas becomes slightly enriched in $^{15}$N, because more \ce{N2} than \mbox{\ion{N}{1}} is frozen out.

\section{Discussion} \label{sec:discussion}
\subsection{Comparisons with observations}
\label{sec:obs}
\begin{figure}[ht!]
\plotone{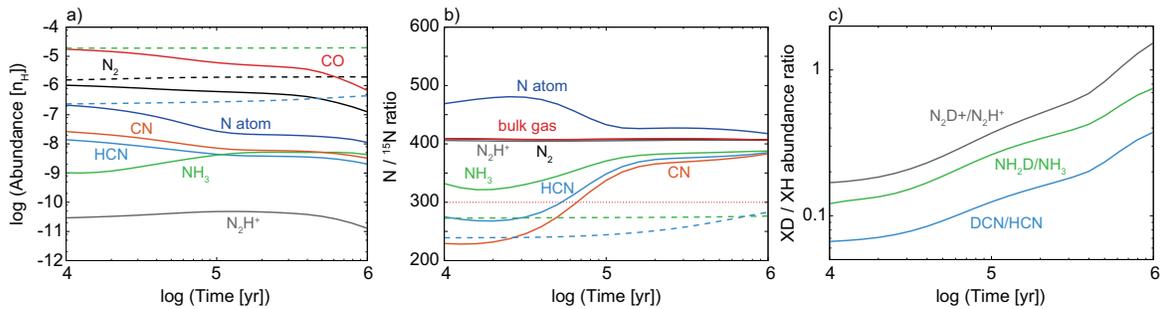}
\caption{
Temporal variations of molecular abundances (panel a), N/$^{15}$N ratios (panel b), 
and abundance ratios between singly-deuterated and non-deuterated species (panel c) in our fiducial model under prestellar conditions.
Solid lines indicate gaseous species, while the dashed lines indicate icy species.
The red dotted line in panel (b) shows [$^{14}$N/$^{15}$N]$_{\rm elem} = 300$.
\label{fig:core}
}
\end{figure}

Table \ref{table:obs} compares the observed and model N/$^{15}$N ratios of \ce{N2H+}, gaseous \ce{NH3}, and nitriles (HCN and CN).
Note that the observed N/$^{15}$N ratios for the nitriles are derived from the observations of the corresponding $^{13}$C isotopologues, 
assuming that the molecular C/$^{13}$C ratios are the same as the elemental C/$^{13}$C ratio.
The observationally derived N/$^{15}$N ratio for the nitriles may be considered as lower limits \citep[see][]{roueff15}.
As a general trend in our models, the N/$^{15}$N ratio decreases in the order, \ce{N2H+}, \ce{NH3}, and the nitriles.
Since our model in the previous section is more appropriate for molecular clouds than for prestellar cores, 
we further run our astrochemical models under prestellar core conditions ($2\times10^5$ cm$^{-3}$, 10 K, and 10 mag) for 10$^{6}$ yr, 
using the molecular abundances at $A_V = 3$ mag as the initial abundances.
Figure \ref{fig:core} shows the temporal evolution of abundances (panel a) and the N/$^{15}$N ratios (panel b) of selected species under the prestellar core conditions 
in the model with $Y_{\rm pd}^{\rm NH_3} = 10^{-3}$ (see also Table \ref{table:obs}).
The abundances of CO and \ce{N2} decrease with time due to freeze-out, while the drop of CO abundance is more significant than that of \ce{N2}.
The \ce{N2H+} abundance is relatively constant, because their formation and destruction rates depend on the abundances of \ce{N2} and CO, respectively.
The gaseous \ce{NH3} abundance increases with time, because the CO freeze-out leads to the enhanced conversion rate from \ce{N2} to \ce{NH3} \citep[e.g.,][]{aikawa05}.
It is confirmed that the $^{15}$N depletion in the bulk gas is almost preserved, 
while the differences in the N/$^{15}$N ratios among the molecules become smaller with time.
The latter point indicates that $^{15}$N fractionation by isotope exchange reactions is not efficient in the prestellar core conditions 
which is consistent with the model of \citet{roueff15}.
Although the rate of cosmic-ray induced photodissociation of N$^{15}$N is higher than that of \ce{N2}, 
this process does not contribute to $^{15}$N fractionation in prestellar cores; 
destruction of \ce{N2} by \ce{He+}. which is the product of cosmic-ray ionization of atomic \ce{He}, 
is much faster than cosmic-ray induced photodissociation of \ce{N2} \citep{heays14}.
Deuterium fractionation, which is driven by hydrogen isotope exchange reactions, becomes more efficient with time under the prestellar core conditions (Figure \ref{fig:core}c).
The different time dependence (or no clear correlation) between $^{15}$N and deuterium fractionation has been reported by \citet{fontani15} and \citet{desimone18}, 
based on the observations of \ce{N2H+}, \ce{N2D+}, and $^{15}$N isotopologues of \ce{N2H+} towards a sample of high-mass and low-mass star-forming cores. 

In the models with $Y_{\rm pd}^{\rm NH_3} \leq 10^{-3}$, all the species in Table \ref{table:obs} are depleted in $^{15}$N, 
which is consistent with the observations in L1544 qualitatively, 
but the models still underestimate the degree of $^{15}$N depletion, in particular, in \ce{N2H+}.
The results presented in this paper are based on a specific physical model.
More numerical studies, varying physical parameters, are necessary to better understand the $^{15}$N observations.

Additional comments on each species are as follows.
\ce{N2H+} is a daughter molecule of \ce{N2}, 
and thus the N/$^{15}$N ratio of \ce{N2H+} follows that of \ce{N2}.
The N$^{15}$NH$^{+}$/$^{15}$NNH$^{+}$ ratio is around unity throughout our simulations, which is consistent with the observations in L1544, but inconsistent with those in B1.
The only known process so far that could deviate the N$^{15}$NH$^{+}$/$^{15}$NNH$^{+}$ ratio from unity 
is the isotope exchange reaction between N$^{15}$N and \ce{N2H+} \citep{adams81,roueff15}.
This pathway would be inefficient in the ISM, since the exchange reaction competes with destruction of \ce{N2H+} by CO, which would be much faster.
Gaseous \ce{NH3} is mainly produced by photodesorption of \ce{NH3} ice with some contribution from the sequential ion-neutral reactions, 
starting from \ce{N2} + \ce{He+}.
Then the N/$^{15}$N ratio of gaseous \ce{NH3} follows that of \mbox{\ion{N}{1}}.
\citet{wirstrom12} found that gaseous \ce{NH3} can be significantly depleted in $^{15}$N (by a factor of around two compared to [N/$^{15}$N]$_{\rm elem}$) 
in some conditions as a result of competition among the three reactions, 
$^{15}$\ce{N+} + \ce{N2}, $^{15}$\ce{N+} + \ce{CO}, and $^{15}$\ce{N+} + \ce{H2}.
The necessary conditions are $200x(\ce{CO}) \lesssim {\rm OPR}(\ce{H2}) \lesssim 60x(\ce{N2})$, 
where $x(i)$ is the abundance of species $i$ with respect to hydrogen nuclei and OPR(\ce{H2}) is the ortho-to-para ratio of \ce{H2}.
In our models, in which adsorption and desorption of \mbox{\ion{N}{1}}, {\ce N2}, and CO are considered,
\ce{N2} is less abundant than CO in the gas phase, even though the binding energy of \ce{N2} is set to be lower than that of CO.
Then the mechanism found by \citet{wirstrom12} does not work in our models.
The critical assumptions in the models of \citet{wirstrom12} and \citet{wirstrom17} are 
that \mbox{\ion{N}{1}} and \ce{N2} do not freeze out onto grain surfaces, while CO does, leading to $x(\ce{N2}) > x(\ce{CO})$ after several freeze-out timescales.
CN and HCN are produced from \mbox{\ion{N}{1}}.
Their N/$^{15}$N ratios basically follow that of \mbox{\ion{N}{1}}, but more enriched in $^{15}$N, due to the isotope exchange reaction between CN and $^{15}$N \citep{roueff15}.

\subsection{$^{15}$N and deuterium fractionation in icy species}
\begin{figure}[ht!]
\plotone{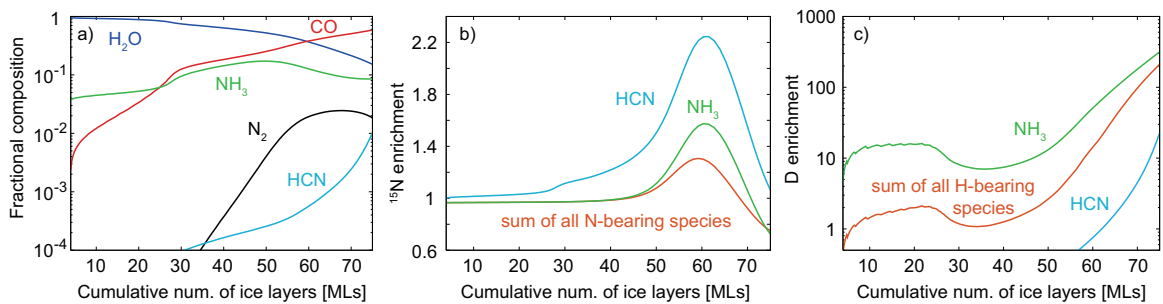}
\caption{Fractional composition, $^{15}$N enrichment with respect to [$^{15}$N/$^{14}$N]$_{\rm elem}$, and deuterium enrichment with respect to [D/H]$_{\rm elem}$ in icy surface species 
as functions of the cumulative number of icy layers formed in the fiducial model.
\label{fig:ice}
}
\end{figure}

Figure \ref{fig:ice} shows the fractional composition (panel a),  the $^{15}$N enrichment (panel b) and the deuterium enrichment (panel c) in \ce{NH3} and HCN ices  
as functions of cumulative number of ice layers formed in our fiducial model.
Around 75 monolayers (MLs) of ice in total are formed by $A_V$ = 3 mag.
\ce{H2O} is the dominant component in the lower ice layers ($\lesssim$60 MLs), while CO becomes more abundant than \ce{H2O} in the upper ice layers.
The CO/\ce{H2O} ratio in the whole ice mantle is 40 \% in our model, which is similar to the the observed median ice composition toward back ground stars \citep[31 \%,][]{oberg11}.
The \ce{NH3}/\ce{H2O} and \ce{HCN}/\ce{H2O} ratios in the whole ice mantle are $\sim$17 \% and $\sim$0.2 \%, respectively, in our fiducial model. 
To the best of our knowledge, no measurements are available for \ce{NH3} ice and HCN ice toward back ground stars in the literature.
The \ce{NH3}/\ce{H2O} ice ratio in our model is higher than the observed median ratio toward 
low-mass and high-mass protostars \citep[][and references therein]{oberg11} by a factor of $\sim$3, 
as often reported in published astrochemical models \citep[e.g.,][]{chang14,furuya15,pauly16}.
If the \ce{NH3} and \ce{NH4+} ice abundances toward the protostellar sources are summed up \citep[see Table 3 of ][]{oberg11}, 
the discrepancy still remains, but becomes smaller (within a factor of 2).

The \ce{NH3} and HCN in the whole ice mantles are enriched both in deuterium and $^{15}$N compared to the elemental ratios, but the degrees of the enrichment are significantly different. 
In our fiducial model, the \ce{NH2D}/\ce{NH3} ratio and the DCN/HCN ratio in the whole ice mantles are $8\times10^{-4}$ and $3\times10^{-4}$, respectively, 
while the N/$^{15}$N ratios of \ce{NH3} ice and HCN ice are 270 and 240, respectively (Table \ref{table:obs}).
The level of $^{15}$N enrichment in the icy species weakly depends on $Y_{\rm pd}^{\rm NH_3}$.
For example, in the model with $Y_{\rm pd}^{\rm NH_3} = 3\times10^{-3}$, the N/$^{15}$N ratios of \ce{NH3} ice and HCN ice in the whole ice mantles are 250 and 190, respectively, 
which are lower than those in the fiducial model.
This trend can be understood, recalling that the maximum level of $^{15}$N enrichment in \mbox{\ion{N}{1}} is 
higher with increasing  $Y_{\rm pd}^{\rm NH_3}$ (Fig. \ref{fig:n_budget}, panels g,h,i) and 
that \ce{N2} ice does not react with other species to form \ce{NH3} ice or \ce{HCN} ice.
Another factor is the timing of \ce{NH3} ice formation; 
in the models with lower $Y_{\rm pd}^{\rm NH_3}$, the amount of \ce{NH3} ice formed prior to the \mbox{\ion{N}{1}}/\ce{N2} chemical transition, which is not fractionated in $^{15}$N, is higher (Fig. \ref{fig:n_budget}, panels a,b,c).
This early formed unfractionated \ce{NH3} ice can overshadow later formed $^{15}$N enriched \ce{NH3} ice.
Then, stronger $^{15}$N depletion in the bulk gas does not necessarily mean stronger $^{15}$N enrichment of icy species in our proposed scenario.
On the other hand, regardless of $Y_{\rm pd}^{\rm NH_3}$, our model predicts that the N/$^{15}$N ratios of icy HCN and icy \ce{NH3} are lower than those of corresponding gaseous molecules.
This prediction can be observationally tested by comparing a molecular N/$^{15}$N ratio in the warm gas ($>$100 K) around protostars, where all ices sublimate, 
with that in the cold outer envelope; the molecular N/$^{15}$N ratio in the warm gas should be lower than that in the cold gas.

The fractionation pattern within the ice mantles is different between deuterium and $^{15}$N.
The concentration of $^{15}$N is nonmonotonic within the ice mantles, while that of deuterim increases from the lower layers to the upper layers.
This is a direct consequence that $^{15}$N fractionation is triggered by isotope selective photodissociation of \ce{N2}, 
while deuterium fractionation is caused by isotope exchange reactions;
the former is most efficient at certain $A_V$, while the latter becomes more efficient with time.
The signatures of the different fractionation mechanisms are preserved in the ice layered structure.
This prediction could be observationally tested by comparing the \ce{NH3}/$^{15}$\ce{NH3} ratio with the \ce{NH2D}/$^{15}$\ce{NH2D} ratio 
in the warm gas ($>$100 K) around protostars, where all ices sublimate;
the \ce{NH3}/$^{15}$\ce{NH3} ratio should be lower than \ce{NH2D}/$^{15}$\ce{NH2D}.
Note that if nitrogen fractionation occurs by isotope exchange reactions as deuterium fractionation, the \ce{NH3}/$^{15}$\ce{NH3} ratio 
should be larger than \ce{NH2D}/$^{15}$\ce{NH2D} \citep[see Fig. 4 of][]{rodgers08b}.

\section{Conclusion} \label{sec:conclusion}
The nitrogen fractionation in the ISM has been a puzzling question.
Previous astrochemical models, which investigate nitrogen isotope fractionation in prestellar cores, have faced difficulties to explain the observations, 
in particular $^{15}$N depletion in \ce{N2H+} in prestellar cores.
In this study, we have proposed that nitrogen fractionation in prestellar cores are largely inherited from their parent clouds, 
which are not fully shielded from the interstellar UV radiation, 
based on our physico-chemical models of molecular clouds.
Around the \mbox{\ion{N}{1}}/\ce{N2} chemical transition,  \mbox{\ion{N}{1}} is enriched in $^{15}$N, while \ce{N2} is depleted in $^{15}$N via isotope selective photodissociation of \ce{N2}.
\mbox{\ion{N}{1}} is adsorbed onto grain surfaces and converted into \ce{NH3} ice by surface reactions, 
while adsorbed \ce{N2} does not react with any other species.
As long as the non-thermal desorption (especially photodesorption in our models) of \ce{NH3} ice is less efficient than that of \ce{N2} ice, 
the net effect is the loss of $^{15}$N from the gas phase, producing the $^{15}$N-enriched ice.
Once the nitrogen isotopes are differentially partitioned between gas and ice in a molecular cloud, 
it should remain in the later stages of star formation (e.g., prestellar core) as long as dust temperature is cold and ice sublimation is inefficient.
If this is the case, $^{15}$N fractionation in dense cores depends on the environments where the cores were formed, rather than the current physical conditions of the cores.
The results presented in this paper are based on a specific physical model.
More numerical studies of $^{15}$N fractionation prior to the formation of dense cores, varying physical parameters, are necessary to better understand the $^{15}$N observations.

The proposed fractionation mechanism is closely related to the conversion of \mbox{\ion{N}{1}} to \ce{N2} in the gas phase.
Regardless of the assumed value of the photodesorption yield of ammonia ice, the \mbox{\ion{N}{1}}/\ce{N2} transition occurs 
before $A_V$ reaches 3 mag in our models,
where one-sided irradiation geometry is employed with the Draine UV radiation field.
Such modest UV attenuation at the transition zone is necessary for the $^{15}$N depletion in the bulk gas,
although it is not easy to observationally constrain when and where the \mbox{\ion{N}{1}}/\ce{N2} transition occurs in the sequence of star formation.
If the depletion of $^{15}$N in the bulk gas is observationally confirmed, it not only verifies our model, but also suggests that 
the \mbox{\ion{N}{1}}/\ce{N2} transition already occurs in the regions, where external UV radiation field is not fully shielded.

\acknowledgments
We are grateful to the anonymous referee for the valuable comments that helped to improve the manuscript. 
This work is partly supported by JSPS KAKENHI Grant Numbers 17K14245 and 16H00093.

\end{document}